\documentclass[10pt,conference]{IEEEtran}
\usepackage{epsfig,rotating,setspace,latexsym,amsmath,epsf,amssymb,amsfonts,bm,bbm,theorem,cite,enumerate,longtable,accents,theoremref,mathtools,mathrsfs,dsfont}
\usepackage{algorithm,graphicx,epsf,authblk,epstopdf,url,color,multirow}
\usepackage{soul,float,algpseudocode}
\usepackage[table,xcdraw]{xcolor}

\newtheorem{theorem}{Theorem}

\newtheorem{lemma}{Lemma}
\newenvironment{Proof}[1]{\medskip\par\noindent{\bf Proof:\,}\,#1}{{\mbox{\,$\blacksquare$}\par}}
\newcounter{example}

\IEEEoverridecommandlockouts
\allowdisplaybreaks

\allowdisplaybreaks

\begin{document}

\title{Dynamic SAFFRON: Disease Control \\ Over Time Via Group Testing}

\author{Batuhan Arasli \qquad Sennur Ulukus\\
	\normalsize Department of Electrical and Computer Engineering\\
	\normalsize University of Maryland, College Park, MD 20742\\
	\normalsize  \emph{barasli@umd.edu} \qquad \emph{ulukus@umd.edu}}

\maketitle

\begin{abstract}
We consider the dynamic infection spread model that is based on the discrete SIR model which assumes infections to be spread over time via infected and non-isolated individuals. In our system, the main objective is not to minimize the number of required tests to identify every infection, but instead, to utilize the available, given testing capacity $T$ at each time instance to efficiently control the infection spread. We introduce and study a novel performance metric, which we coin as $\epsilon$-disease control time. This metric can be used to measure how fast a given algorithm can control the spread of a disease. We characterize the performance of dynamic individual testing algorithm and introduce a novel dynamic SAFFRON based group testing algorithm. We present theoretical results and implement the proposed algorithms to compare their performances.  
\end{abstract}

\section{Introduction}
The group testing approach is proposed by Dorfman in \cite{dorfman1943}. Its main objective is to identify the infection status of a group of individuals by performing fewer tests than the number of individuals. The main idea behind group testing is to pool the obtained test samples and to test these pooled samples, instead of testing each sample individually. By this way, when a mixed sample results negative, it implies that every test sample included in that mixed sample is negative. On the other hand, if a mixed sample results positive, it implies that there is at least one positive test sample among the test samples that are mixed in that mixed sample. The main objective of the group testing problem is to design the testing scheme by specifying the test pools such that the infected set is identified with a minimum number of tests.

Since the group testing scheme that was proposed by Dorfman in \cite{dorfman1943}, various group testing algorithms have been studied. Mainly, two classical infection models have been considered in majority of works: combinatorial model where the number of infections is fixed and the infected set is uniformly randomly selected from all fixed sized subsets of all individuals, and probabilistic model where each individual is infected with a fixed probability, independently. The group testing algorithms can be divided into several categories: adaptive group testing algorithms where the tests are performed in multiple steps and test results from previous steps can be used while designing the future test pools and non-adaptive group testing algorithms where all of the tests are designed to be performed in a single step. The group testing algorithms can be categorized depending on their decoding error performance where zero-error group testing algorithms identify the infected set perfectly and small (vanishing) error algorithms identify the infected set with a small error that vanishes with large population. As a variation of the test result model, noisy test results setting has been proposed and studied. For these classical settings, various algorithms have been proposed, analyzed and the capacities have been characterized for different parameter regimes \cite{binarysplittingorig,hwang_binary,hwang_disjunct,RUSZINKO,nonadaptive_bounds,adaptivecapacity,mazumdar_nonadaptive,scarlett_noisynonadaptive,wu_partition,wang_combquant,combinatorial_gt,ddintroduction,sharper,scarlettbook,cai_noisy,karimi_irregularsparsegraph,inan_optimalityks,johnson_nearconstant,allemann,atia_saligrama_first}.

Group testing works that are based on the classical settings have a limitation for the overall performance: the advantage of group testing diminishes as the prevalence rate of the infection increases, due to the nature of the problem \cite{RUSZINKO,scarlettbook}. In more recent works, the classical settings have been challenged: non-identical infection probabilities, correlations between infection status of the individuals and community structure based available side information scenarios have been proposed and studied \cite{prior_stats,doger2021group,correlated_bio,lincorrelated,diggavicommunity,diggavioverlap,ayferozgurcommunity,arasli2021group,edgefaulty2022,structidalino,generalsetsys,noisysideinfo}. These practical settings prove that by challenging the classical settings, further improvements can be obtained: group testing approach can be used efficiently even when the prevalence rate is high among the population if further available side information is used in test designs. However, the overall static nature of the group testing problem is not challenged in these works. For a given group of individuals, the aim is to identify the infected set, as in the classical settings. In a recent line of works, the static nature of the majority of group testing works has been challenged and dynamic models are proposed \cite{acemoglu2021optimal,dynamicgtdiggavi,dynamicgtdiggavi2,doger2022dynamical,arasli2022dynamic}. In these works, SIR \cite{sirmodel2020} based dynamic infection spread models are studied. Testing and identification are done over time while the infection status of the individuals also evolve.

We consider the discrete time, SIR based dynamic infection spread model that we introduced in \cite{arasli2022dynamic}. In our model, at each time instance, the testing capacity is limited and fixed, and full identification of the infections in the system is not possible. Each discrete time instance consists of two phases: infection spread phase and testing phase. Identified infections are isolated and further infection spread by them is prevented. Since the testing capacity is limited at each time instance, not all of the infections are detected and infected individuals that are not detected and isolated keep spreading the infection. The dynamic infection spread model is based on the SIR model, where the population is divided into three disjoint groups: susceptible individuals (S), non-isolated infections (I), and isolated infections (R). At the initial time instance, we assume i.i.d.~initial infections in the system and at each time instance that follows, non-isolated infections spread the disease to the susceptible individuals in the system. Consecutively, during the testing phase, limited number of tests are performed and detected infections are isolated. We assume that, for the rest of the testing process, isolated infections remain in the same state. This is a reasonable assumption since they eventually recover, and even when their isolation ends, they can be immune to the infection for a disease-specific time period. We assume that the testing process finalizes prior to the end of that time period. In our model, in contrast to the prior works, the objective is not to identify all of the infections while minimizing the number of tests, but to identify as many infections as possible at each time instance by using limited number of $T$ tests and to eventually control the spread of the disease.

In this paper, we expand the framework that we introduced in \cite{arasli2022dynamic}. We consider the disease control time, $\bar{t}$ and $\epsilon$-disease control time $\bar{t}_\epsilon$ as novel performance metrics in our analysis. We characterize the performance of dynamic individual testing algorithm for these novel performance metrics. Moreover, we introduce a novel dynamic algorithm: \emph{dynamic SAFFRON based group testing algorithm} which is inspired by the static SAFFRON scheme that is studied in \cite{aldridgeisolate} and introduced in \cite{saffron}. We present the mean-sense theoretical analysis of both dynamic individual testing algorithm and dynamic SAFFRON based group testing algorithm. Finally, we implement the discrete time SIR based dynamic infection spread model and the dynamic algorithms to simulate them and compare the numerical results with the theoretical results that we obtain.

\section{System Model} \label{sec2}
There are $n$ individuals in the system and their infection status, which we denote by $U_i(t)$ for individual $i$ at time $t$, changes over discrete time instances. At any given time, there are three disjoint groups in the population: susceptible individuals who can get infected (S), non-isolated infections who have not been detected (I) and isolated infections who have been detected by performed tests and isolated indefinitely (R). $U_i(t)$ can take values 0 (susceptible), 1 (non-isolated infection) and 2 (isolated infection), representing the infection state of the individual $i$ at time $t$. At $t=0$, we introduce the initial i.i.d.~infections to the system: each individual is infected with probability $p$, independently. We introduce three random processes: $\alpha(t)$ denotes the number of susceptible individuals, $\lambda(t)$ denotes the number of non-isolated infections and $\gamma(t)$ denotes the isolated infections, at time $t$.

Each discrete time instance $t\geq 1$, consists of two consecutive phases: infection spread phase and testing phase. During the infection spread phase, each non-isolated and infected individual can infect each susceptible individual with probability $q$, independently. Here, since each susceptible individual can be infected by each non-isolated and infected individual, this can be modelled as the realization of $\alpha(t)\lambda(t)$ independent Bernoulli random variables with parameter $q$, for each time $t \geq 1$. Following the infection spread phase, the testing phase starts in which the testing capacity is limited to a given, fixed number, $T$. Thus, the aim of designing the group testing algorithm is to construct $T \times n$ binary test matrices that specify the testing pools. Let $\bm{X}(t)$ denote the binary test matrix for the tests at time $t$, and $\bm{X}_{ij}(t)$ denote the $i$th row and $j$th column of $\bm{X}(t)$, where $\bm{X}_{ij}(t)$ is 1 if the test sample of $j$th individual is mixed in the $i$th mixed test sample, and 0 otherwise. Let $y(t)$ denote the test result vector of the tests performed at time $t$ and $y_i(t)$ denote the $i$th test result at time $t$. Then, we have
\begin{align}
	y_i(t) = \bigvee _ {j \in [n]} \bm{X}_{ij}(t) \mathds{1}_{\{U_{j}(t)=1\}}, \quad i \in [T]
\end{align}

We assume that results of the tests that are performed at time $t$ are available before the infection spread phase at time $t+1$, i.e., $y(t)$ vector is known while designing the binary test matrix $\bm{X}(t+1)$. Moreover, when infected individuals are detected during the testing phase at time $t$, they are immediately isolated and thus, prevented to spread the infection for the rest of the testing process, i.e., $U_i(t) = 2$ for all times $t \geq t'$, if the individual $i$ is detected to be infected by the tests performed at time $t'$. Note that, the only possible infection state changes in our system is either from susceptible individuals to non-isolated infections, which happens via infection spread from non-isolated and infected individuals to susceptible individuals during the infection spread phase at each time, or from non-isolated infections to isolated infections which happens via infection detection during the testing phase at each time instance. A group testing policy $\pi$ is a scheme that specifies the algorithm to construct binary test matrices $\bm{X}(t)$ for each time instance $t \geq 1$, until the infection is under control.

Since the source of the infection state evolution in our dynamic model is the non-isolated infections, we define the disease control time $\bar{t}$ to be the first time when there are no non-isolated infections remaining in the population, i.e., $\bar{t} = \min \{t ~|~ \lambda(t) = 0\}$. After time $\bar{t}$, every individual in the population is either susceptible or isolated, i.e., $U_i(t) = 0$ or $U_i(t) = 2$ for $i \in [n]$ and for all $t \geq \bar{t}$. Furthermore, we introduce $\epsilon$-disease control time for probabilistic analysis, denoted by $\bar{t}_\epsilon$ where $E[\lambda(\bar{t}_\epsilon)] = \epsilon$ holds. While characterizing and analyzing the performance of a group testing algorithm $\pi$, there are two performance metrics associated with it: the disease control time $\bar{t}$ (or $\bar{t}_\epsilon$) and the number of individuals that have never gotten infected throughout the process, i.e., $\alpha(\bar{t})$. Lower $\bar{t}$ is favored to control the disease faster, and higher $\alpha(\bar{t})$ is favored to control the disease with lower number of total infections. 

\section{Proposed Algorithms and Analysis} \label{sec3}
In this section, we introduce a novel dynamic group testing algorithm: dynamic SAFFRON based group testing algorithm. This algorithm is inspired by the static group testing algorithm introduced in \cite{saffron}, coined as SAFFRON scheme, which is further studied in \cite{aldridgeisolate} for partial detection problem. We analyze dynamic SAFFRON based group testing algorithm in terms of two performance metrics that we consider: disease control time and number of susceptible individuals when the disease is brought under control. Furthermore, we analyze the dynamic individual testing algorithm that we introduced in \cite{arasli2022dynamic}, in terms of its $\epsilon$-disease control time performance. 

\subsection{Related Prior Results}
For the sake of completeness, here we present a revised version of the results from \cite{arasli2022dynamic}, that we use for the analysis in this paper. To keep it compact, we only include the main results that we use in the analysis in this paper.

We consider \emph{symmetric and converging dynamic testing algorithms} which must satisfy the \emph{symmetry} criterion:
\begin{align}
    P(U_i(t)=k) = P(U_j(t)=k) , \quad k \in \{0,1,2\} \label{symalgcr}
\end{align}
for each time instance $t \geq 0$ and for all $i,j \in [n]$ pair. Moreover, they must satisfy the \emph{convergence} criterion:
\begin{align}
    \lim_{t \rightarrow \infty}P(U_i(t)=1) = o(1/n), \quad i \in [n] \label{convalgcr}
\end{align}
Let $p'(t)$ denote the probability of an individual not being identified during the testing phase at time $t$. Note that, since we consider symmetric and converging dynamic testing algorithms, this probability is the same for all individuals. We consider dynamic testing algorithms where $p'(t)$ only depends on the testing capacity $T$, $\alpha(t)$, $\lambda(t)$ and $\gamma(t)$. We assume the infection spread probability $q$ scales as $o(1/n)$. Since the infection spread is realized for every susceptible and non-isolated infected individual pair and there can be $O(n^2)$ such pairs at each time instance, assuming $q$ to be $o(1/n)$ is a mild assumption. 

In the remainder of this subsection, we present our prior result from \cite{arasli2022dynamic} that we use in our analysis in this paper, regarding symmetric and converging dynamic testing algorithms.

We start with the following lemma, where we prove that symmetric and converging dynamic testing algorithms guarantee the disease to be controlled eventually.

\begin{lemma} \label{lemma1}
When a symmetric and converging dynamic testing algorithm is implemented, $\lim_{t \rightarrow \infty}E[\lambda (t)] = o(1)$ and thus, the system approaches the steady state, in the mean sense.
\end{lemma}

The statements of Lemma~\ref{lemma2} and Theorem~\ref{theorem1} give improved characterizations of the approximation terms compared to the original statements in \cite{arasli2022dynamic}. The approximations are characterized by $o(1)$ terms in this paper, which follow from the linear approximations in Taylor series expansions for exponential terms and hold since $q= o(1/n)$. Moreover, arbitrarily small variance conditions in \cite{arasli2022dynamic} are also characterized to scale as $o(1)$ here as well. In the following lemma, we characterize an approximation of the probability of an individual being infected and non-isolated at time $t$, i.e., probability of the event of $U_i(t) = 1$. 

\begin{lemma} \label{lemma2}
When a symmetric and converging dynamic testing algorithm is implemented, we have
\begin{align}
    P(U_i (t) = 1) = p((1+nq(1-p)))^t \prod \limits_{j=1}^{t}p'(j) + o(1)
\end{align}
where the conditional probability of an individual not being identified in the tests at time $t$ given $\lambda (t-1)$ is denoted by $p'_{\lambda (t-1)}$ and when $cov\left(P(U_i(t)=0|\lambda (t)),p'_{\lambda (t)}(t+1)\right)$ and $cov\left(P(U_i(t)=1|\lambda (t)),p'_{\lambda (t)}(t+1)\right)$ both scale as $o(1)$ with respect to $n$ for all $t \geq 0$.
\end{lemma}

In the following theorem, we present a general result that characterizes the approximation of the expectation of $\alpha(t)$. This result holds for all symmetric and converging dynamic testing algorithms that satisfy the small covariance conditions that we state in Lemma~\ref{lemma2}.

\begin{theorem} \label{theorem1}
When a symmetric and converging dynamic testing algorithm is implemented and vanishing covariance constraints in Lemma~\ref{lemma2} are satisfied for all $t \geq 0$, we have
\begin{align}
    E[\alpha (t)] = n(1-p)(1-q)^{np\sum \limits_{i = 0}^{t-1}\left((1+nq(1-p))^i \prod \limits _{j=1}^{i}p'(j)\right)} + o(1) \label{thm1}
\end{align}
\end{theorem}

We use Theorem~\ref{theorem1} to characterize the expected number of susceptible individuals when the disease is brought under control, which is one of the two performance metrics that we consider. On the other hand, for the characterization of the disease control time $\bar{t}$, our analysis is based on Lemma~\ref{lemma2}.

\subsection{Dynamic Individual Testing Algorithm}
In dynamic individual testing algorithm, we consider randomized, individual testing of $T$ individuals at each time instance $t \geq 1$, where $T$ individuals to be tested are uniformly randomly selected from whole population, independent across both individuals and time. In \cite{arasli2022dynamic}, we proved that dynamic individual testing algorithm (weak dynamic individual testing algorithm in that paper) is a symmetric and converging dynamic testing algorithm. Furthermore, we derived $E[\alpha(t)]$ results for the disease control time. Here, we derive $\epsilon$-disease control time performance results of dynamic individual testing algorithm.

In the following theorem, we present our result for the $\epsilon$-disease control time performance metric when dynamic individual testing algorithm is used in our proposed dynamic system.

\begin{theorem} \label{theorem2}
When dynamic individual testing algorithm is used, we have
\begin{align}
    \bar{t}_\epsilon = \frac{\ln (\epsilon/np + o(1))}{\ln \left((1-\frac{T}{n})(1+nq(1-p))\right)}
\end{align}
\end{theorem}

\begin{Proof}
We start with the mean of non-isolated infections
\begin{align}
    E[\lambda(t)] &= n P (U_i(t) = 1) \label{thm2l1}\\
    &=np((1+nq(1-p)))^t \prod \limits_{j=1}^{t}p'(j) + o(n) \label{thm2l2}
\end{align}
where \eqref{thm2l1} follows from the definition of $\lambda(t)$ as the total number of non-isolated infected individuals and \eqref{thm2l2} follows from Lemma~\ref{lemma2}. Then, we have
\begin{align}
    \epsilon &= np((1+nq(1-p)))^{\bar{t}_\epsilon} \prod \limits_{j=1}^{\bar{t}_\epsilon}p'(j) + o(n) \label{thm2l3}\\
    &=np\left((1-\frac{T}{n})(1+nq(1-p))\right)^{\bar{t}_\epsilon} + o(n) \label{thm2l4}
\end{align}
where \eqref{thm2l3} follows from the definition of $\epsilon$-disease control time and \eqref{thm2l4} follows from the fact that $p'(t) = (1-T/n)$ for all $t \geq 1$ when dynamic individual testing algorithm is used. Finally, we get
\begin{align}
    \bar{t}_\epsilon &= \frac{\ln (\epsilon/np + o(1))}{\ln \left((1-\frac{T}{n})(1+nq(1-p))\right)}
\end{align}
by arranging the terms. 
\end{Proof}

In Theorem~\ref{theorem2} we characterize the $\epsilon$-disease control time metric for dynamic individual testing algorithm. Since it is defined as the time instance when there are $\epsilon$ non-isolated infected individuals in the mean sense, $\epsilon$-disease control time metric presents a characterization of how fast a given algorithm can control the disease spread.

\subsection{Dynamic SAFFRON Based Group Testing Algorithm}
We propose and analyze a novel algorithm, which we coin as dynamic SAFFRON based group testing algorithm. It is inspired by the static group testing algorithm, SAFFRON, introduced in \cite{saffron} and studied in \cite{aldridgeisolate} for partial detection.

The static SAFFRON scheme inspired algorithm presented in \cite{aldridgeisolate} is based on the idea of constructing binary test matrices for the groups of individuals that contain approximately one infected individual. Constructed SAFFRON scheme test matrices can be used to recover exactly one infected individual within the tested group, while also indicating whether there is zero or more than one infected individuals within the tested group if there is not exactly one infected individual.

Here, we propose the novel dynamic SAFFRON based group testing algorithm, where at each time instance $t+1$, groups of size $\lfloor(n-\gamma(t))/E[\lambda(t)]\rfloor$ are selected uniformly and randomly from the set of non-isolated individuals. For the $\lambda(t)$ values that are close to their mean,\footnote{Our numerical results suggest a concentration around the mean for $\lambda(t)$. An in-depth concentration analysis can be the subject of future works.} each of these groups has approximately one infected and non-isolated individual.

By designing SAFFRON scheme inspired binary test sub-matrices that are guaranteed to identify one infection from each group that has exactly one infection, we detect and isolate infections over time.

For a selected group of individuals the binary test sub-matrix is constructed as follows:
\begin{itemize}
    \item Let $\eta$ denote the size of the selected individuals, i.e., $\eta = \lfloor(n-\gamma(t))/E[\lambda(t)]\rfloor$.
    \item First $\lceil\log (\eta )\rceil$ rows of the test sub-matrix are constructed where column vector that corresponds to column $i$ is set to the binary representation of the number $i-1$.
    \item Then, remaining $\lceil\log (\eta )\rceil$ rows of the test sub-matrix is constructed by concatenating the created sub-matrix in the previous step with the binary matrix where the value of each element is flipped, i.e., XORed with 1.
\end{itemize}
By using this construction, it is guaranteed that, if there is exactly one infection within the tested individuals, there will be exactly $\lceil \log (\eta) \rceil$ positive tests and the positive test indices will be the binary representation of $i-1$ where the infected individual index is $i$. If there is no infection, then all tests will be negative. In the case of more than one infection, strictly more than $\lceil \log (\eta) \rceil$ tests will be positive. Thus, this construction guarantees the detection of a single infection within the selected group.

In the following lemma, we characterize the expected number of detected infections at each time instance.

\begin{lemma}\label{lemma3}
When dynamic SAFFRON based group testing algorithm is used, at each time instance $t \geq 1$, in average
\begin{align}
    \frac{T}{2}\log ^{-1} \left( \frac{n-\gamma(t-1)}{E[\lambda (t-1)]}\right)\left(1-\frac{E[\lambda(t-1)]}{n-\gamma (t-1)}\right)^{\frac{n-\gamma (t-1)}{E[\lambda (t-1)]}-1}
\end{align}
infections are detected and isolated.
\end{lemma}

\begin{Proof}
We choose the group size as $\frac{n-\gamma(t-1)}{E[\lambda (t-1)]}$ since there will be approximately one infected individual within each group when the groups are chosen uniformly randomly from the set of all non-isolated individuals. Since the testing capacity is $T$ at each time instance, there will be 
\begin{align}
    \frac{T}{2}\log ^{-1} \left( \frac{n-\gamma(t-1)}{E[\lambda (t-1)]}\right)
\end{align}
such groups. Recall that the probability that a non-isolated individual is infected is $\frac{E[\lambda(t-1)]}{n-\gamma (t-1)}$ due to the fact that dynamic SAFFRON based group testing algorithm uniformly randomly selects the groups at each time instance and since the infection statistics are symmetric over individuals. Furthermore, the group size is inverse of this term. Therefore, each group contains exactly one infected individual with probability
\begin{align}
\left(1-\frac{E[\lambda(t-1)]}{n-\gamma (t-1)}\right)^{\frac{n-\gamma (t-1)}{E[\lambda (t-1)]}-1}
\end{align}
where lemma statement follows.
\end{Proof}

Similar to our analysis for dynamic individual testing algorithm, we can use the results of Lemma~\ref{lemma2} and Theorem~\ref{theorem1} for dynamic SAFFRON based group testing algorithm where we have 
\begin{align}
    p'(t)=\frac{\zeta}{E[\lambda(t)]}
\end{align}
where $\zeta=    \frac{T}{2}\log ^{-1} \left( \frac{n-\gamma(t-1)}{E[\lambda (t-1)]}\right)\left(1-\frac{E[\lambda(t-1)]}{n-\gamma (t-1)}\right)^{\frac{n-\gamma (t-1)}{E[\lambda (t-1)]}-1}$. To justify using  Lemma~\ref{lemma2} and Theorem~\ref{theorem1} results, we need to prove that dynamic SAFFRON based group testing algorithm satisfies symmetry and convergence criteria in \eqref{symalgcr} and \eqref{convalgcr}.

Since the selection of tested groups is independent across individuals for each time instance, dynamic SAFFRON based group testing algorithm is symmetric. To guarantee convergence, we consider using dynamic individual testing algorithm whenever $T<2\log\lfloor(n-\gamma(t))/E[\lambda(t)]\rfloor$, i.e., the regime where expected number of non-isolated infections is small with respect to the total number of non-isolated individuals. In that regime, SAFFRON based construction cannot be used efficiently to detect infections. In the regime where expected number of non-isolated infections is high, SAFFRON based scheme can consistently detect infections as characterized in Lemma~\ref{lemma3}.  

\begin{figure}[t]
	\centering
	\epsfig{file=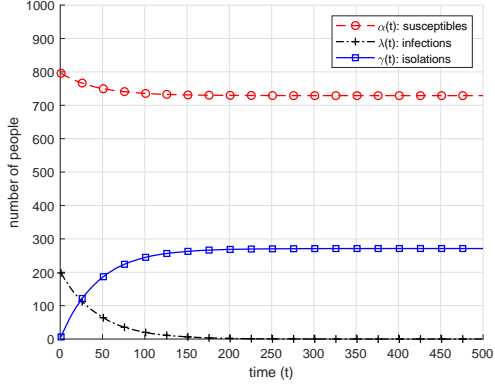,width=.4\textwidth}
	\caption{Numerical averages of the random processes $\alpha(t)$, $\lambda(t)$ and $\gamma(t)$ for the system parameters $n=1000$, $T=30$, $q=0.00001$, $p=0.2$, for dynamic individual testing algorithm.}
	\label{fig1}
	\vspace*{-0.4cm}
\end{figure}

\section{Numerical Results}

In our simulations, we implement the dynamic discrete SIR based infection spread system. We initialize our system with random independent initial infections with probability $p$ and in following time instances, we simulate the random disease spread from non-isolated infections to susceptible individuals, independently with probability $q$. For both dynamic algorithms, we run our system for 500 time instances and iterate realizing this dynamic system for 1000 times. We save the sample path trajectories of the random processes $\alpha(t)$, $\lambda(t)$ and $\gamma(t)$, take their averages and plot. 

\begin{figure}[t]
	\centering
	\epsfig{file=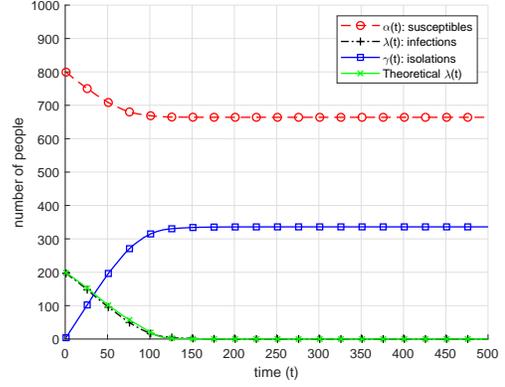,width=.4\textwidth}
	\caption{Numerical averages of the random processes $\alpha(t)$, $\lambda(t)$ and $\gamma(t)$, with theoretical calculation of $\lambda(t)$, for the system parameters $n=1000$, $T=30$, $q=0.00001$, $p=0.2$, for dynamic SAFFRON based group testing algorithm.}
	\label{fig2}
  \vspace*{-0.4cm}
\end{figure}

To control disease spread, in our first simulation, we implement dynamic individual testing algorithm, where randomly selected $T$ individuals are individually tested at each time instance. Here, by using our theoretical result in Theorem~\ref{theorem2}, we have $\bar{t}_\epsilon=235.57$ when $\epsilon = 1$. This is consistent with the numerical results plotted in Figure~\ref{fig1}. In our second simulation, we implement a hybrid version of dynamic SAFFRON based group testing algorithm. At each time instance, we randomly select groups of people where each group has one infection in average, where we use expected $\lambda(t)$ that we calculate by using Lemma~\ref{lemma3}, to obtain the group size. Then, we construct SAFFRON scheme based test sub-matrices for each of these groups and perform tests, with maximum $T$ tests in total. In practice, SAFFRON scheme based construction can have non-utilized testing capacity at each discrete time instance due to the fact that the testing capacity $T$ may not be divisible by the calculated group size. We further utilize these available testing capacity and perform random individual testing when there are available non-utilized tests remain. Moreover, later in the testing process, since the expected $\lambda(t)$ becomes smaller, calculated group size grows larger and required number of tests for SAFFRON based construction exceeds the testing capacity $T$. For these later times in the testing process, we switch to the dynamic individual testing algorithm to detect and isolate remaining infections. We also plot theoretical calculation of expected $\lambda(t)$ in Figure~\ref{fig2}, where we use Lemma~\ref{lemma3} for calculation, which align with the numerical results that we obtain, as we plot in Figure~\ref{fig2}. We observe that the steady state mean number of susceptible individuals is slightly higher in dynamic individual testing algorithm while the convergence time is slightly lower in dynamic SAFFRON based group testing algorithm. This observation for this set of parameters suggest that both algorithms can be used to optimize different performance metrics, depending on the system requirements and parameters, as well as hybrid usage of the dynamic algorithms is also possible as in our hybrid dynamic SAFFRON based group testing algorithm implementation.

\section{Conclusion}
We considered a dynamic infection spread model based on the well-known discrete SIR model. We aimed to efficiently utilize the available testing capacity $T$ at each time instance to control disease spread within a community rather than minimize the number of required tests to determine every infection as in the static group testing problem. Furthermore, we presented a novel performance metric: $\epsilon$-disease control time. Different than our previous metric in \cite{arasli2022dynamic}, that is the expected number of susceptible individuals in the steady state, $\epsilon$-disease control time measures the timeliness of the disease control. We introduced \emph{dynamic SAFFRON based group testing algorithm} and presented mean-sense performance results, which can be further used in combination with our previous results regarding symmetric and converging dynamic testing algorithms. 

\bibliographystyle{unsrt}
\bibliography{references_grouptesting}

\end{document}